\documentclass[draft]{agujournal2018}
\usepackage{url} 
\usepackage{lineno}
\usepackage[english]{babel}
\usepackage{amsmath}
\usepackage{epsfig}	
\usepackage{psfig}		
\usepackage{setspace}
\usepackage{graphicx,color}		
\usepackage{times}
\usepackage{tabularx}
\usepackage[bottom]{footmisc}
\usepackage{tablefootnote}

\draftfalse
\journalname{Earth and Space Science}

\begin{document}
\justify
\title{Time-Latitude Distribution of  Prominences for 10 Solar Cycles: A study using Kodaikanal, Meudon and Kanzelhohe Data}

\authors{Subhamoy Chatterjee\affil{1}, Manjunath Hegde\affil{1}, Dipankar Banerjee\affil{1,2}, B. Ravindra\affil{1}, Scott W. McIntosh\affil{3}}

\affiliation{1}{Indian Institute of Astrophysics, Koramangala, Bangalore 560034, India}
\affiliation{2}{Center of Excellence in Space Sciences India, IISER Kolkata, Mohanpur 741246, West Bengal, India}
\affiliation{3}{High Altitude Observatory, National Center for Atmospheric Research, PO Box
3000, Boulder, Colorado 80307, USA}

\correspondingauthor{Dipankar Banerjee}{dipu@iiap.res.in}

\begin{keypoints}
\item Calibration of Ca~{\sc ii}~K disc-blocked digitised images as recorded in Kodaikanal observatory for the period of 1906-2002.
\item Automated detection of prominences for 10 solar cycles (1906-2018) from Ca~{\sc ii}~K disc-blocked digitised images as recorded in Kodaikanal Observatory and H$_\alpha$ images from Meudon and Kanzelhohe Observatory.
\item Clear identification of pole-ward migration for all the cycles, estimation of migration rates from pole-ward branches near the onset and close to pole through piece-wise linear fits with error bars. Those can act as vital inputs to polar field build-up process. 
\end{keypoints}

\begin{abstract}
Solar prominences are structures of importance because of their role in polar field reversal. We study the  long-term variation of the time latitude distribution of solar prominences in this article. To accomplish this, we primarily used the digitised disc-blocked Ca~{\sc ii}~K spectroheliograms as recorded from Kodaikanal Solar Observatory for the period of 1906 -- 2002.  For improving the data statistics we included full disc H$_{\alpha}$ images from Meudon and Kanzelhohe Observatory  which are available after 1980. We developed an automated technique to identify the latitudinal locations of prominences in daily images from all three datasets.  Derived time-latitude distribution clearly depicted pole-ward migration of prominence structures for 10 cycles (15-24). Unlike previous studies, we separated the rate of pole-ward migration during on-set and near pole, using piece-wise linear fits. In most cases, we found acceleration in pole-ward migration with the change occurring near $\pm 70^\circ$ latitudes. The derived migration rates for such large number of solar cycles can provide important inputs towards understanding polar field build-up process.




\end{abstract}

\section{Introduction}
Magnetic field is responsible for the heterogeneity as  observed on the Sun in different temporal and spatial scales. 
Continuous monitoring of the solar magnetic field has been possible only in recent past with the launch of space-borne telescopes like Michelson Doppler Imager (MDI, on the Solar Heliospheric Observatory, SoHO, \citep{1997SPD....28.0127H}) and more recently Helioseismic and Magnetic Imager (HMI, on the Solar Dynamics Observatory, SDO \citep{2012AAS...22020701F}). Before that magnetic field measurements in low resolution was initiated by ground-based observatories like Mount Wilson and most prominently by Wilcox Solar Observatory since the year 1976 \citep{{1992AAS...180.5106B},{2000eaa..bookE4331.}} and KittPeak/National Solar Observatory(NSO) since 1970 \citep{{1971IAUS...43...51L},{2011CoSka..41..106M}}. These data sources have been extensively  used for solar cycle studies \citep{2015TESS....111102H}. To trace back the solar magnetic field activity before 1976 different proxies have been used. Some of the prominent features serving as proxies are  sunspots, plages and filaments/prominences.
	
Prominences/filaments are structures formed in the solar atmosphere following magnetic polarity inversion lines. They are formed in the chromosphere by cool dense (hundred times cooler and  denser than the coronal material) plasma held in place by solar magnetic fields \citep{{Engvold2015},{2018LRSP...15....7G},{Martin2015}}. At the limb they appear as bright features when observed in few optical, EUV lines such as H$_\alpha$, Ca~{\sc ii}~K, He {\sc ii} 304 \AA. They appear as dark elongated hairlike features on the disc for example in H$_\alpha$ 6562.8 \AA~or in He {\sc ii} 304 \AA ~(named as filaments).  Prominences present themselves in different morphology, lifetime and complexity in magnetic field environments. Orientation of small leg-like structures called `barbs' bifurcating from filaments, known as `chirality', provides important information about the supporting magnetic field morphology. \citet{2001ApJ...558..872M} presented an interesting model on formation and evolution of filaments/prominences of different chiralities through `head-tail linkage' of different bipolar magnetic regions resulting in `convergence and cancellation'. Another comprehensive study on filament formation and its hemispheric distribution was made by \citet{Mackay2015}. \citet{2009SoPh..254...77Y} connected global magnetic field distribution and origin of hemisphere-wise filament chiralities. As a follow-up study, \citet{2012ApJ...753L..34Y} predicted dependance of high-latitude filament hemispheric chirality patterns on cycle phase using NSO/Kitt Peak cycle 23 magnetograms and emphasized on importance of coronal magnetic field memory. Thus long term study of prominences over several cycles and at different phases of solar cycle can give valuable insight on the physics of the solar atmosphere.  There have been several studies to correlate eruptive prominence distributions and coronal mass ejections (CMEs) \citep{{2000ApJ...537..503G},{2003ApJ...586..562G},{2008ApJ...674..586S},{Gopalswamy2015},{Webb2015},{Fan2015},{Lugaz2015}}. \citet{{1908Natur..78..174L},{1922MNRAS..82..323L},{1931MNRAS..91..797L}} using coronal drawings and photographs obtained during the total solar eclipse suggested that there is an intimate connection between the distribution of prominences around the solar limb and the forms of corona. Though prominences form all over the Sun, their latitudinal distribution changes significantly with time being correlated with global properties of large scale magnetic fields on solar surface. 

As pointed out earlier, prominences were first seen during total solar eclipses. Early history of prominence observations can be found in \citet{{1974GAM....12.....T},{1998ASPC..150...11T}}. Zone of polar prominences and its pole-ward migration was discovered by \citet{1872MNRAS..32..226S}. To probe the change in time-latitude distribution of prominences and its heterogeneity for the understanding and predicting solar magnetic fields over longer time scales \citep{2018NatCo...9.5209B}, one needs long-term, homogeneous datasets. Surface flux transport dynamo models have largely been effective in realizing of polar field build-up which remains to be the governing factor in predicting solar cycles \citep{{2018NatCo...9.5209B},{2012ApJ...753..146M}}. It is thus worth understanding the long-term behaviour of polar prominences and their connection to polar magnetic fields. Along with the observations of on-disc filaments with spectroheliograph, systematic prominence observations  were also carried out. Daily prominence observation at the Kodaikanal Solar Observatory (KoSO hereafter) started before the observational set-up for H$_\alpha$ was ready. This is because of the fact that the prominences were observed in Ca~{\sc ii}~K at KoSO above the solar limb by blocking the disc. So, full-disc and disc blocked observation started simultaneously at KoSO in Ca~{\sc ii}~K wavelength using spectroheliograph. Meudon observatory started prominence observation in H$_\alpha$ consistently after KoSO since 1919. Lomnicky Peak Observatory also recorded solar prominences in H$_\alpha$ from 1967 \citep{{1988CoSka..17...63R},{1990SoPh..128..253B},{1994CoSkS..24..135R},{1998SoPh..177..357M}}  until 2009. It is worth mentioning that Kislovodsk solar station in Russia started observing the Sun in H$_\alpha$ since 1957 and recorded filaments, prominences in the form of synoptic charts \citep{{2001SoPh..198..409M},{2001SoPh..202...11M},{2003SoPh..214...41M}}. The observatory continues to make this observations till today. Recently, ground based disc-blocked prominence observation in H$_\alpha$ has been initiated at Kanzelhoehe Solar Observatory since 2009.

 There have been several works reported in past using these historical data on prominences. From such datasets both pole-ward and equator-ward migration were shown by \citet{1957sun..book.....A}. \citet{1973SoPh...28..389W} established 3 narrow zones, which show different latitudinal behaviour in the 11-year cycle, namely sunspot prominences, stable long-lived prominences and polar zone prominences forming at latitudes around 45$^\circ$ during minimum and migrating towards poles around solar maximum gaining a velocity of 10 -- 25 ms$^{-1}$. The change of migration rates from lower latitudes to near poles was pointed out by \citet{2001SoPh..202...11M}. \citet{azamb} made a large number of measurements on filaments and prominences during the period of 1919 -- 1930 and found on an average that all prominences have a low pole-ward drift in both the hemispheres. \citet{2000SoPh..194...87V} studied the distribution and asymmetry of Solar Active Prominences (SAP) for the period of 1957 -- 1998 (solar cycles 19 -- 23) and found that  E -- W asymmetry of SAP events is not significant. He reported that N -- S asymmetry of SAP events is significant and it has no relation with the solar maximum or solar minimum during solar cycles. Another study on asymmetry of SAP spatial distribution was made by \citet{2009SoPh..260..451J}. \citet{2012ApJ...744..168L} using  Solar  LImb  Prominence  CAtcher  and Tracker (SLIPCAT), studied prominences during the period 2007 -- 2009. \citet{2013PASJ...65S..16S}  using Nobeyama Radioheliograph reported the unusual migration of prominences producing region of activities in the southern hemisphere and interpreted the anomalies from the distribution of the photospheric magnetic field during the cycles 23 -- 24. \citet{2014LRSP...11....1P}  reviewed characteristic properties of prominences and filaments as derived from observations. Other studies were made to quantify the time-latitude distribution of prominence \citep{{1990SoPh..128..253B},{1998SoPh..177..357M}} and correlated them with other solar cycle indices \citep{2001SoPh..202...11M} for making cycle prediction \citep{2003SoPh..214...41M}. Long-term study on filaments have also been made using hand-drawn Synoptic maps from Meudon Observatory \citep{2016SoPh..291.1115T} and MacIntosh database \citep{2018ApJ...868...52M}. These studies provided important results on filament tilt angles, their latitude dependence, rush to the poles and polar reversal. 

Most of the aforementioned  works focused  on temporal variation in spatial distribution of filaments/prominences but are limited to either short time-span or hand-drawn data. Naturally, more reliable results require a uniform digitised long-term dataset in gray-scale. Combined data from different observatories require cross calibration, otherwise detection techniques with same parameters but as applied to different data are not comparable. \citet{2011CoSka..41..133R} cross-calibrated data from  Lomnicky Peak Observatory
(1967 -- 2009)  and Kanzelhoehe Solar Observatory (2009 -- 2010) before detection of prominences. In this context, it is worth mentioning that KoSO recorded prominence observations from 1906 until 2002 on daily basis with an instrument of unchanged optics. This ensured uniformity in the KoSO prominence data quality making it advantageous for feature detection. Prominence data from Kodaikanal for the period 1904 -- 1914 is discussed by \citet{evershed} and by \citet{moss} for the period 1905 -- 1928. Half-yearly summaries of prominence observations were published in the Bulletin of the Kodaikanal Observatory. Using this data, prominence eruption was also reported in \citet{1917KodOB...3..209E}. But the notable study from KoSO prominence data was made by \citet{1952Natur.170..156A}. He performed manual identification of prominences during the period 1905 -- 1950 and found a relationship between prominence activity and sunspot cycle. General review and discussion about the prominence data collected at Kodaikanal observatory was summarised by \citet{Ananthakrishnan1954}. KoSO photographic plates have been recently digitised for the entire duration of observations. In modern times it is important to identify and detect structures such as prominences using computer codes for avoiding human bias and increasing speed. \citet{2011CoSka..41..133R} have used the digitized Lomnicky Peak Observatory and Kanzelhohe Solar Observatory data  to detect the prominence structures through semi-automated codes. Later, they have cross-calibrated these two datasets as well.  

Our objective is to detect prominence locations from the digitized KoSO data in a fully automated manner. In the present study to improve the statistics and to make the detection up-to-date, we also included the full disc H$_{\alpha}$ data from Meudon (1980-2002) and Kanzelhohe Observatory (2000-2018) by blocking the disc and bringing them to equal footing as KoSO data.
In this article we present an automated technique  to identify the prominence latitudes from combined (KoSO-Meudon-kanzelhohe) disc-blocked dataset.  We describe the data in Section~\ref{ddes} followed by analysis of data in Section~\ref{ada} encompassing calibration, prominence detection technique and results on pole-ward migration. We finally discuss the results of the study in Section~\ref{Disc} and conclude in Section~\ref{concl}. 

\vspace{.1\textwidth}
\section{Data Description}\label{ddes}
In following three subsections we describe the three datasets used in our study for prominence detection.
\subsection{Kodaikanal Dataset}
The Ca~{\sc ii}~K line is very broad in wavelength coverage with 3 prominent positions namely K$_1$, K$_2$ and K$_3$ on the line profile. The K$_3$ region (3933.7\AA), is where we see filaments and prominences similar to those seen in H$_\alpha$ but with much less contrast and less well defined boundaries. Photographic observations of solar prominences in Ca~{\sc ii}~K line (central wavelength 3933.7\AA, passband 0.5\AA) using spectroheliograph were started at Kodaikanal during 1905 \citep{Ananthakrishnan1954}. The nominal spectral passband remained constant at 0.5\AA~throughout the study time \citep{2017ApJ...841...70C}. However, there are studies that predict change of bandwidth and center wavelength for this dataset from change in contrast of the images \citep{{2019A&A...625A..69C},{2019arXiv190805493C}}.With weather permitting (when the sky and seeing conditions were good), the routine Ca~{\sc ii}~K observations were made between 02:00 to 04:30 UT  at KoSO with the spectroheliograph. This data is available in 8-bit and 16-bit digitised form. The 8-bit version was presented in \cite{Ermolli_2009} as the most uniform dataset in terms of resolution, large-scale inhomogeneity and contrast among different available long-term datasets. An accurate photometric calibration of this series was made by \cite{2018A&A...609A..92C} to make the dataset ready for irradiance calculation and also detection of plages \citep{2019A&A...625A..69C}. Recently, the 16-bit series was studied by \cite{2019arXiv190805493C} confirming its superiority over the 8-bit series in terms of uniformity.  It is the same instrument which was used to make the disc-blocked Ca~{\sc ii}~K spectroheliograms as well. In order to get the contrast for the prominences, the chromosphere (disc region) was mechanically blocked by a circular mask of same size before capturing the image on photographic plates/films. The telescope used for this had an optics with a 30 cm objective, with f-ratio of f/21 \citep{2014SoPh..289..137P}. This optics was unchanged throughout the observation period ensuring uniformity in quality. The disc blocked images were digitized recently through a 16-bit digitization unit at KoSO consisting of an uniform illumination source, imaging optics for proper magnification and a  $4096 \times 4096$ CCD cooled at $-100^\circ$C. The digitized images  from 1906 until 2002 (Table~\ref{data_source}) used in our study were in all `.fits' format and of size $4096 \times 4096$ with a pixel scale of $\approx$ 0.88 arcsec.  Figure~\ref{rawim}a shows the typical digitized disc-occulted chromospheric image. The date and time of the observation are written on each of the plate at one corner, on the opposite side of emulsion material. The white line on the right bottom corner is the holder for occulter disc.  Double and single dots outside the occulter represent the North and South pole of the Sun respectively.
Due to the unavailability of pole marking on the images taken before 1906, we used here the data starting from 1906. 
 Many prominences could be seen in the digitized image on the limb.  Figure~\ref{statim}a depicts the number of days of observational data available per year for different years. On an average $\approx 269$ observing days are available each year for the analysis before 1975. But, after 1975 the average number of days of available data per year drops to $\approx 143$. Other data sources, as mentioned in Table~\ref{data_source}, were used to complement Kodaikanal data in later years  and complete the analysis till current solar cycle.

\subsection{Meudon Dataset}
Digitised full-disc H$_\alpha$ spectroheliograms (Figure~\ref{rawim}b) from Meudon Observatory \citep{1985LAstr..99..557D} starting from the year 1980 till May, 2002 were used in our study (Table~\ref{data_source}). The image sizes varied from $928 \times 942$ to $1007 \times 1003$. Yearly histogram of observing days is depicted in Figure~\ref{statim}b. The average number of images available per year over the study period for Meudon is $\approx254$.

\subsection{Kanzelhohe Dataset}
\begin{figure*}[!htbp]
\centering
\includegraphics[scale=0.23]{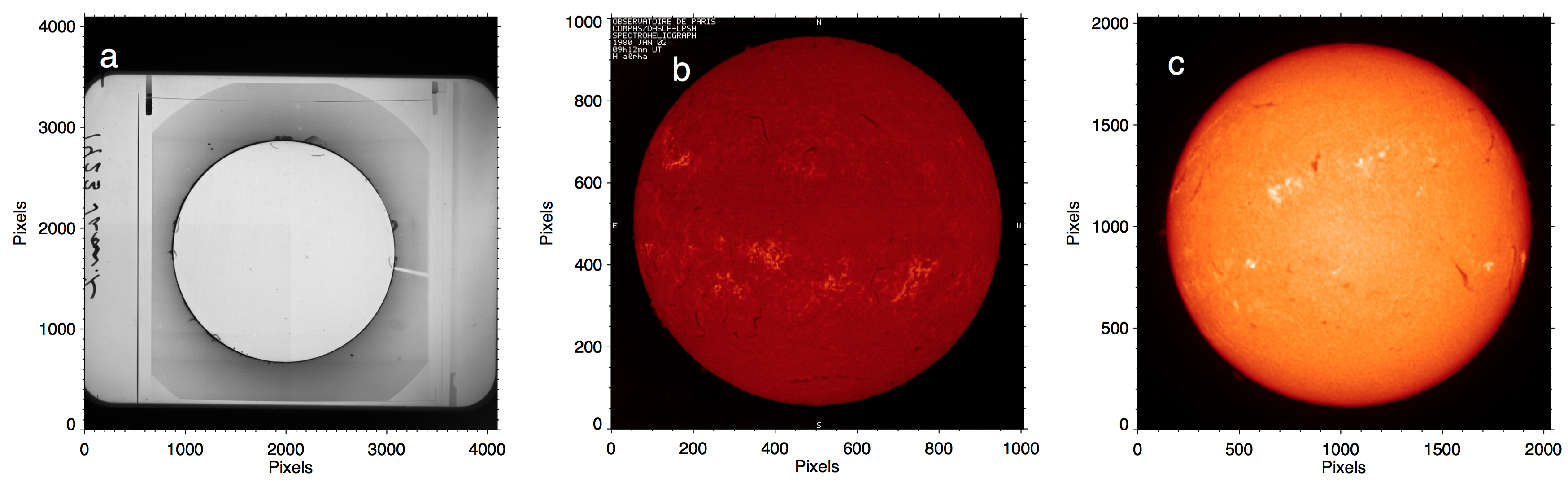}
  \caption{Raw datasets used for prominence detection. a) Representative raw KoSO Ca~{\sc ii}~K disc blocked spectroheliogram captured on January 02, 1923; b)  Representative raw Meudon H$_\alpha$ full disc spectroheliogram captured on January 02, 1980; c) Representative raw Kanzelhohe H$_\alpha$ full disc filtergram captured on January 21, 2003.}
 \label{rawim}
\end{figure*}

\begin{figure*}[!htbp]
\centering
\includegraphics[scale=0.7]{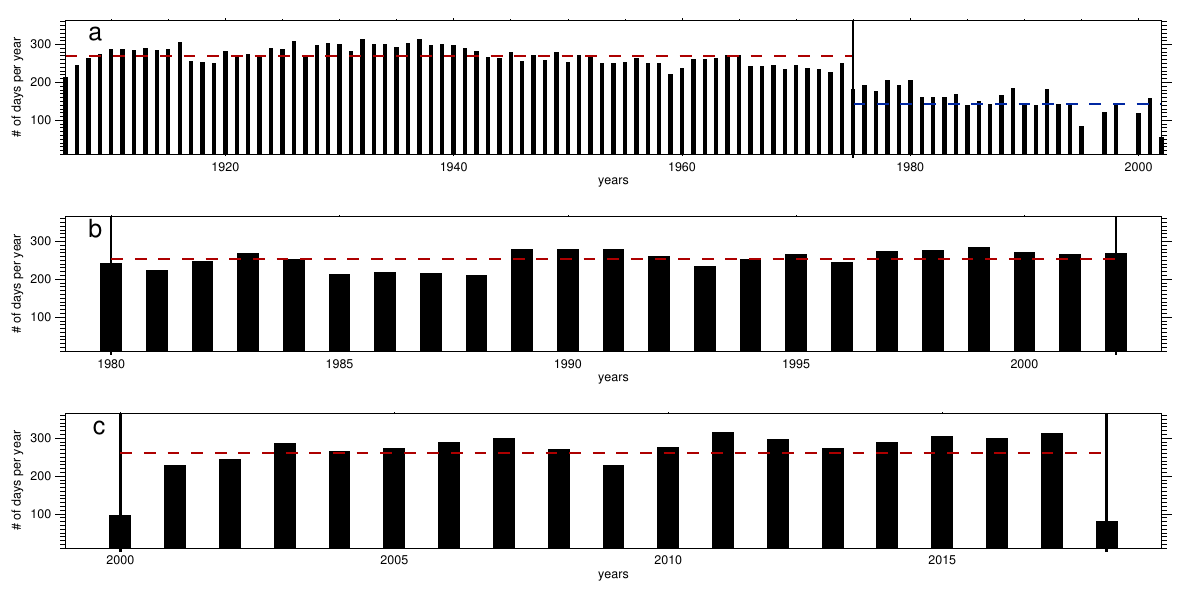}
  \caption{Histograms of observing  days during the time period studied for KoSO, Meudon and KSO. a) Yearly histogram of available days in KoSO Ca~{\sc ii}~K disc blocked dataset. The red and blue dashed lines indicate the yearly mean before and after 1975 respectively; b)  Yearly histogram of available days in Meudon H$_\alpha$ full disc dataset.  The red dashed line indicates the mean yearly observing days for entire study period; c) Yearly histogram of available days in KSO H$_\alpha$ full disc dataset.  The red dashed line indicates the mean yearly observing days for entire study period.}
 \label{statim}
\end{figure*}

Kanzelhohe Solar observatory (KSO) full-disc H$_\alpha$ images (Figure~\ref{rawim}c) starting from the year 2000 till April, 2018 were used in this study (Table~\ref{data_source}). KSO comes under Global H$_\alpha$ Network \citep{2013EGUGA..15.1459P}. These filtergrams were captured through two different CCDs before and after the year 2008. Because of that, the images are of size $2032\times2032$ (before 2008) and $2048\times2048$ (since mid January, 2008). Yearly histogram of observing days is depicted in Figure~\ref{statim}c. The average number of images available per year over the study duration for KSO is $\approx260$.  

\begin{table}
\begin{center}
\caption{Datasets used for prominence detection}\label{data_source}
\begin{tabular}{ c  c  c }
\hline
Observatory  & Data Type &  Time span \\	   \hline  
 	   \hline  
 	  Kodaikanal & Ca~{\sc ii}~K disc blocked & 1906--2002\\  \hline
    Meudon & H$_\alpha$ full disc & 1980--2002\tablefootnote{\url{http://bass2000.obspm.fr/data_guide.php}}\\   \hline                  
Kanzelhohe & H$_\alpha$ full disc & 2000--2018\tablefootnote{\url{http://cesar.kso.ac.at/}}\\\hline
\end{tabular}
\end{center}
\end{table}

\begin{figure*}[!htbp]
\centering
\includegraphics[scale=0.23]{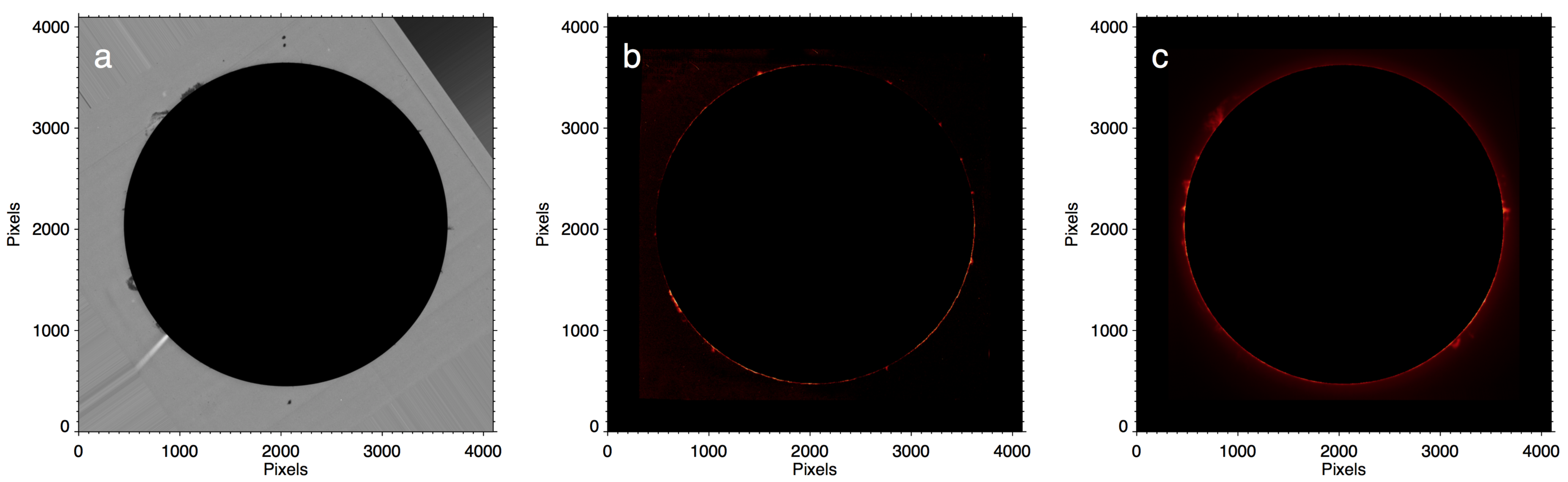}
  \caption{Calibrated disc-blocked images. a) Disc centered and radius normalised version of Figure~\ref{rawim}a; b) Disc blocked, centered and radius normalised version of Figure~\ref{rawim}b; c) Disc blocked, centered and radius normalised version of Figure~\ref{rawim}c.}
 \label{calim}
\end{figure*}

\section{Analysis of Data}\label{ada}
\subsection{Calibration}
Before detection of prominence locations, the digitised images had to go through few calibration steps. In the following two subsections we introduce the steps to calibrate both Kodaikanal Ca~{\sc ii}~K disc-blocked images and Meudon, Kanzelhohe H$_{\alpha}$ full-disc images in detail to make them ready for automated detection of prominence location using a single algorithm.
\subsubsection{Calibration of Kodaikanal Ca~{\sc ii}~K disc-blocked images}
All the digitized $4096\times4096$ raw images were  initially binned to a size of $512\times512$ and `Prewitt' \citep{sonka2014image} edge detection operator  was applied on the same to highlight  the solar disc edge. Subsequently the edge detected image was thresholded to convert it to a binary form. The resulting binary image showed different edge features, prominent in the photographic film. To separate the disc out of  all the features in the image, we used Circle Hough Transform \citep{{sonka2014image},{2016ApJ...827...87C}}. For reducing the time and complexity we imparted a small search range about a nominal radius. This nominal radius value in pixel was obtained by dividing the disc angular size in arcsec, varying over a year, with the image pixel scale in arcsec. Subsequently, we acquired values of disc center coordinates and its radius. 

To find the large scale intensity variation outside the solar disc, resized $512\times 512$ images were blurred using 2D median filter of size $15\times15$ pixels. This filter size was a trade-off between prominence dimension and edge effect. The filtered image was subsequently resized to original size of $4096\times4096$. This method has been commonly used to estimate asymmetric center-to-limb variation of intensity in full disc solar Ca~{\sc ii}~K images \citep{{2010SoPh..264...31B},{2016ApJ...827...87C}}. The raw image was thereafter divided with the median filtered image to generate a normalized image.

It should be noted that just median filtering is not an accurate method for background normalization when the objective is to make the data ready for irradiance study \citep{2018A&A...609A..92C} and may affect detection of features, measurement of contrast. We did not need the data to be photometrically calibrated as we opted for only detection of prominence locations avoiding measurement of individual feature areas. 

Using the information obtained from Hough transform, we generated disc centered normalized images with a margin of $30\%$ of the disc radii ($r$, in pixels) on both sides of solar disc to have provision of detecting large prominences. All the disc centered images of size $(2\lfloor{1.3r}\rceil+1)\times(2\lfloor{1.3r}\rceil+1)$ ($\lfloor{.}\rceil$ stands for the nearest integer function or the rounding operation) 
were converted to images of size $4096\times4096$. This step made disc radii constant throughout the study time. Because of the change of Sun-Earth distance over a year the radial extent of the images ($1.3r$ or 2048 pixels) varies between 1226 arcsec and 1265 arcsec. Also, in terms of image scale this variation is within $\approx$ 0.6 -- 0.62 arcsec/pixel. It can be seen from Figure~\ref{rawim}a that the raw images have north-south pole marking denoted by double  and single dots outside solar disc for north and south poles respectively. These dots were placed manually on the photographic plates/films considering p-angle astronomical ephemerides and rotation of images due to instrument over a day. They ensure alignment of images post digitization as small errors may occur while placing the plates inside digitizer unit. By detecting those dots we found the angle by which the image has to be rotated so that the North pole will be located right in upward direction. The region within a radius of $1600$ pixels was removed to get rid of any effect due to any misalignment of blocking disc in the instrument and the actual solar disc. It is worth mentioning that the misalignment causes a stretch of solar limb to be connected to the actual prominences causing over-detection. A representative calibrated image from the raw image depicted in Figure~\ref{rawim}a is shown in Figure~\ref{calim}a.

\subsubsection{Calibration of Meudon and Kanzelhohe H$_{\alpha}$ images}
Meudon Observatory images despite having N-S marks in vertical directions were not always aligned. Each of them were checked and aligned using p-angles from astronomical ephemerides as and when required. Full-disc images from Meudon and Kanzelhohe observatory (Figures~\ref{rawim}b,c) were first smoothed using median filter and subsequently edge detection was performed to detect the disc edges. It is to be noted that apart from yearly variation of disc size, we dealt with different image sizes from both Meudon and Kanzelhohe for different periods of time.  So, the size of the smoothing kernels was varied accordingly.  On the binary image containing limb, Circle Hough transform was applied to find the disc center and radius. Using those information, disc centering was performed and the centered image was rotated for p-angle compensation.
The disc portion of centered and rotation corrected image was then blocked using a circular mask. The disc-blocked image depicts the prominences (Figures~\ref{calim}b,c). Before applying the detection algorithm all these disc blocked  images were brought to the size of $4096\times4096$ (same as Kodaikanal) keeping $30\%$ margin outside limb.
%

\subsection{Detection of Prominence locations}
\begin{figure}
\centerline{
\includegraphics[scale=1.0]{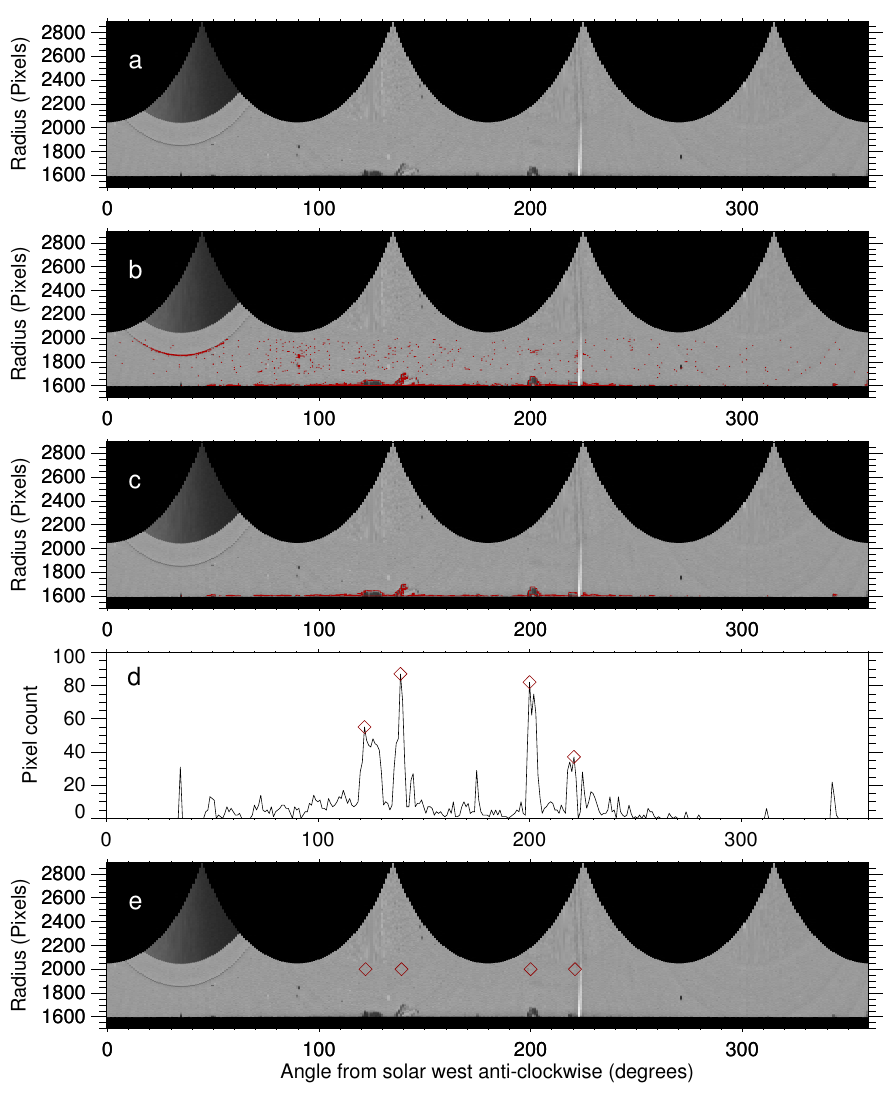}}
\caption{Steps for detection of prominence locations. a) Polar map generated from Figure~\ref{calim}a;  b) Red contours  correspond to the segmented regions after intensity thresholding of the polar map; c) Contours obtained after removing regions not connected to limb from (b); d) Counts of detected pixels  along radius (c) for each angle w.r.t. solar west is plotted and local maxima with higher counts in red symbols are considered to be locations of prominence structures; c) Prominence locations identified in (d) over plotted on (a) with red symbols. As the solar disc does not contain any information here, a part of it near limb has been shown as black strip at the bottom of every panel.}
\label{det}
\end{figure}

The calibrated images ($I_{cal}$), described in the last section, with image-center coordinates $(c_x, c_y)$ were first converted to polar form ($I_{polar}$) using eq.(\ref{eq}).


   \begin{align}
    I_{polar}(R,\theta)=
    \begin{cases}
      I_{cal}(\lfloor{c_x+R\cos\theta}\rceil, \lfloor{c_y+R\sin\theta}\rceil)
,&  \text{if}\ \lfloor{c_x+|R\cos\theta|}\rceil<4096 \\ & \text{and}~ \lfloor{c_y+|R\sin\theta|}\rceil<4096 \\
      0 , &  \text{otherwise}
    \end{cases}
     \label{eq}
       \end{align}
       
with $c_x=2048$, $c_y=2048$, $0\le R< \lfloor{2048\sqrt{2}}\rceil$ and $0^\circ\le\theta<360^\circ$. Rounding operation ($\lfloor{.}\rceil$) has been used to get integral pixel coordinates.
Subsequently, polar images of size $2896$ pixels $\times360$ pixels (Figure~\ref{det}) were produced from disc centered calibrated images of size $4096\times4096$ (Figure~\ref{calim}). For the polar maps pixel scale values obtained along $R$ and $\theta$ are $\approx$ 0.6 -- 0.62 arcsec and 1$^\circ$ respectively.

The periodic curved patterns in the polar image $I_{pol}$ stand for the four straight sides of $I_{cal}$ and the regions above the curved portion are dark due to no data outside square $I_{cal}$. It can be seen that the curved edges in $I_{pol}$ starts appearing for $R>2048$ for obvious reasons. So, for segregating the prominences we considered $I_{pol}$ pixels only for $1600<R<2000$. Within this region we calculated median (med$_{pol}$) and standard deviation ($\sigma_{pol}$) of pixel intensity. Thereafter we produced binary image ($BW_{polar}$) applying a threshold of (med$_{pol}$-$\sigma_{pol}$) to KoSO polar images making all the pixel intensity below that threshold as 1 and rest as 0. For Meudon and Kanzelhohe polar images, binary version was produced by making all pixel intensities above (med$_{pol}$+$\sigma_{pol}$) as 1 and rest as 0. This difference was introduced considering whether prominence grey level is brighter or dimmer compared to background. Figure~\ref{det}b depicts the binary image contours overplotted in red on the polar image shown in Figure~\ref{det}a. This prominence segmented binary image presents itself with several problems. Firstly, it captures small scale intensity fluctuations in the background. Secondly, near the limb sudden jump of intensity occurs as a result of median filtering which connects the prominence structures making disjoint regions to appear as single connected region. Finally, the thresholding captures curved artefacts which are manifestations of scratches in the original images. As the background small scale non-uniformity in intensity has no relation with $\theta$, count of thresholded pixels along $R$ should follow the prominence height fluctuations. To partially get rid of background non-solar features, all the detected regions not connected to solar limb were removed. This step is demonstrated through Figures~\ref{det}b,c. From the resulting binary image with reduced noise, we generated prominence radial pixel count curves as function of $\theta$ expressed by-$$c(\theta)=\sum_{R} BW_{polar}(R,\theta)$$ $c(\theta)$ corresponding to Figure~\ref{det}c is shown in Figure~\ref{det}d. 

It was obvious that local maxima of prominence heights are representatives of prominence locations. So, we calculated the locations of local maxima $\{\theta_M\}$ of $c(\theta)$. This process can be described as,
$$\{\theta_M\}=\underset{0^\circ\le\theta <360^\circ}{\operatorname{arg}} \{\frac{dc(\theta)}{d\theta}=0 ~\textrm{and}~ \frac{d^{2n}c(\theta)}{d\theta^{2n}}<0\}$$
We refined these maxima locations by putting a  threshold as function of mean ($m$) and standard deviation ($\sigma$) over $c(\theta_M)$. Finally we described the prominence locations as \[\Theta=\theta_M(\underset{k}{\operatorname{arg}} {\{c(\theta_M(k))>m+0.7\sigma\}})\] The coefficient of $\sigma$ was selected by several iterations to select prominence locations efficiently. Count maxima at the locations $\Theta$ corresponding to Figure~\ref{det}a are marked with red symbols on the count curve $c(\theta)$ in Figure~\ref{det}b. Including the $B_{\circ}$ angle correction, $\Theta$ was converted to prominence latitudes ($L_{promin}$) in the range $[-90^\circ, 90^\circ]$ using eq.(\ref{eq2}).

\begin{equation}
    L_{promin}=\sin^{-1}(\cos(B_{\circ})\sin(\Theta))
     \label{eq2}
       \end{equation}
       
Along with the latitudes we also recorded the counts i.e. $c(\Theta)$.   Figure~\ref{det}c depicts the prominence locations with red symbols plotted above the prominences. It can be observed that the north and south pole markings did not get detected as prominences because they are detached from the limb. 

\begin{figure}
\centerline{
\includegraphics[scale=0.7]{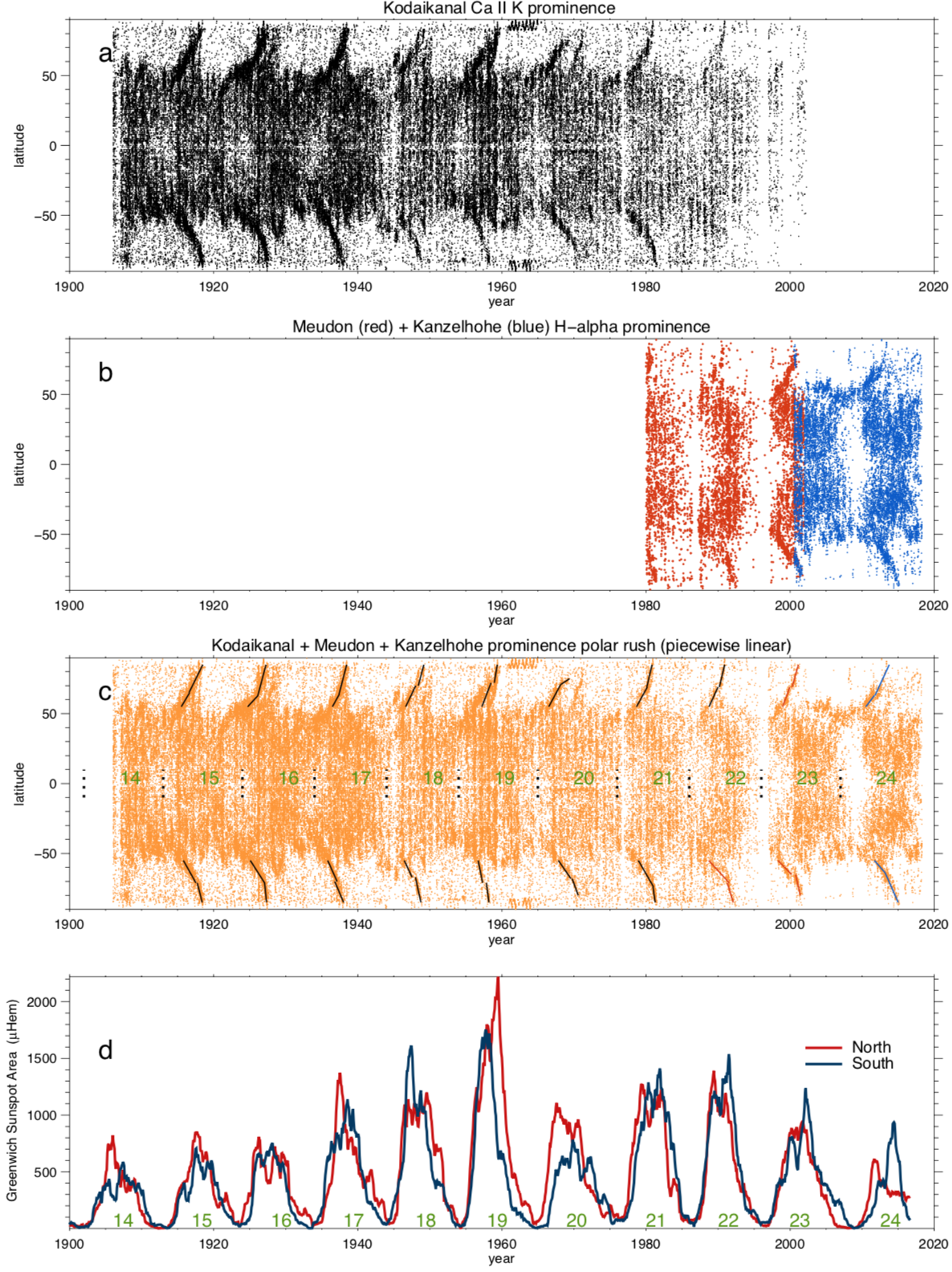}}
\vspace{-0.02\textwidth}
\caption{Prominence locations from 1906 until April, 2018.  a) Time-latitude distribution of the KoSO prominences clearly depicting polar rush till latitudes close to $90^\circ$ for cycles 15-21 in both northern and southern hemispheres;  b) Detected prominence locations using same algorithm on Meudon and Kanzelhohe H$_{\alpha}$ data.; c) Piece-wise linear fits to the polar branches depicting the latitudes from which migration rate changes. Black, red and blue lines are generated respectively from KoSO, Meudon and Kanzelhohe data. Combined time-latitude distribution of prominences have been put in background with orange dots; d) 13-month running average smoothed Greenwich sunspot area cycle till 2016 for north and south.}  
\label{pole_rush}
\end{figure}

From the detection of prominence locations, we obtained time latitude diagrams for prominences for Kodaikanal, Meudon and Kanzelhohe Observatory. Figure~\ref{pole_rush} depicts time-latitude distribution and fitted polar branches of prominences from 1906 till April, 2018. Time-latitude pattern of prominences detected using KoSO dataset is shown in Figure~\ref{pole_rush}a. From this diagram we immediately notice the clear signature of pole-ward migration till the close vicinity of poles from  cycle 15 through cycle 21 both in northern and southern hemispheres. Cycle 20 depicts two northern polar branches with the second being weaker than the first one. The nature of polar rush from cycle 15 until cycle 18 can be directly compared with the time-latitude plots depicted in \citet{{1952Natur.170..156A},{Ananthakrishnan1954}}. Polar branch remains clear for northern hemisphere in cycle 22 whereas that in the southern hemisphere do not show clear trend. This is the effect of reduced number of observing days after cycle 21. The same algorithm applied on Meudon and Kanzelhohe H$_{\alpha}$ images manages to provide clear depiction of pole-ward migration from cycle 22 through cycle 24 (till mid of cycle 23 for Meudon in red and mid cycle 23 till cycle 24 for Kanzelhohe in Blue) in both hemispheres (Figure~\ref{pole_rush}b). The combined dataset is presented with orange symbols in Figure~\ref{pole_rush}c. Figure~\ref{pole_rush}d depicts the 13-month running average of monthly sunspot area variation in north and south from Greenwich dataset \footnote{\url{http://solarscience.msfc.nasa.gov/greenwch.shtml}}.
Also, near minima there is consistent reduction in prominence counts at low latitudes for all the cycles. Cycle 23/24 minima shows a clear depiction of this and also that the prominences are created $\approx 55^\circ$ latitude (Figure ~\ref{pole_rush}b).



\subsection{Polar Rush Fitting}
Before fitting polar branches of prominence time-latitude diagram, we concatenated the three datasets namely KoSO, Meudon and Kanzelhohe. Subsequently, 2-dimensional histogram of prominence time-latitude was generated using latitude bin of $1^\circ$ and time bin of 0.1 years. Using the histogram density values the time-latitude plot was converted to an image. That image was thresholded to generate the binary version. For every latitudinal step, the mean temporal locations and standard error of the same were found for all the polar branches. It should be noted that we used each polar branch from one of the three observatories having major contribution in terms of data points. Now, most of earlier studies, calculated pole-ward migration rates through single linear fit. However, we clearly notice acceleration of polar rush in the time-latitude plot from our detection (Figures~\ref{pole_rush}a,b). This has been reported earlier, indicating the change of polar rush close to pole \citep{2001SoPh..202...11M}, though no concrete method of such estimation was presented. Fitting nonlinear functions such as higher order polynomials suffer from latitudinal dependence of error bars. To avoid this and also to quantify the change in migration rate with time ($t$), we opted for fitting with a piece-wise linear function defined by eq.(\ref{eq3}).


     
\begin{align}
   t=
    \begin{cases}
      p_0^{(1)}+p_1^{(1)}\theta
, & \text{if} ~~55^\circ \leq \theta < \theta_{sep}~ (\text{Low Polar Zone})\\
      p_0^{(2)}+p_1^{(2)}\theta
      , & \text{if} ~~\theta \geq  \theta_{sep}~ (\text{High Polar Zone })\\
    \end{cases}
     \label{eq3}
       \end{align}

The separator latitude $ \theta_{sep}$ was varied such that the fitting error is minimised. In this case two zones of pole-ward migration are created with rates $1/p_1^{(1)}, 1/p_1^{(2)}$. The rate uncertainties for those are given by $\Delta p_1^{(1)}/(p_1^{(1)})^2, \Delta p_1^{(2)}/(p_1^{(2)})^2$. Uncertainties in the coefficients denoted by $\Delta p_1^{(1)}, \Delta p_1^{(2)}$ were found by supplying the standard errors of mean temporal locations while fitting. \\Hereafter we use LPZ and HPZ as abbreviations for `Low Polar Zone' and `High Polar Zone' respectively. 

Polar rush fits on the combined dataset using piece-wise linear function are marked by  black lines in Figure~\ref{pole_rush}c. 
As pointed out in several earlier studies \citep{{Ananthakrishnan1954},{1990SoPh..128..253B},{1998SoPh..177..357M}}, after sunspot minimum prominence activity begins to
develop in the high latitude zones between $40^{\circ} - 50^{\circ}$ latitudes and  with the progress in sunspot activity, the prominence activity shifts towards pole with a rapid rush near sunspot maximum. Very similar phenomena can be noted from in the time-latitude plots (Figures~\ref{pole_rush}a-d) when compared with epochs sunspot area cycle maxima (Figures~\ref{pole_rush}e) for both the hemispheres. The mean of absolute differences between epoch of sunspot maxima and epoch of prominences reaching pole over all cycles comes about $\approx 1.5$ years in north and $\approx 0.8$ year in south from our detection. 

\begin{figure}[!thbp]
\centerline{
\includegraphics[scale=0.6]{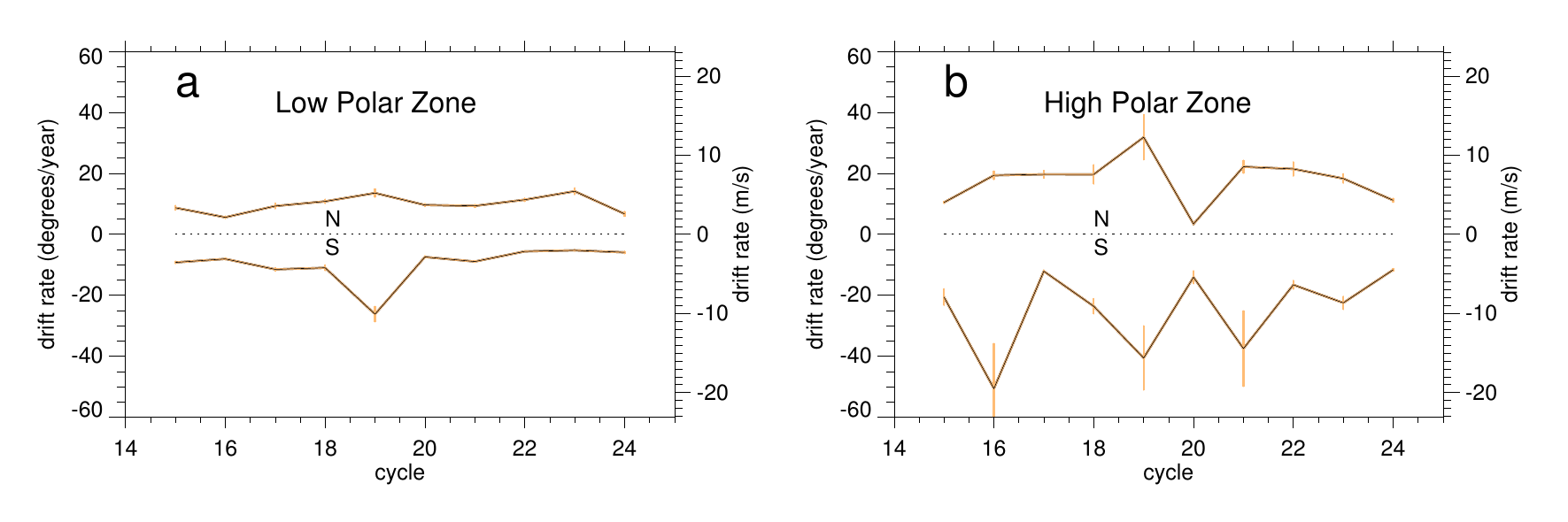}}
\caption{Cycle wise drift rate of polar prominences with error bars as vertical stretches in two latitude zones extracted from piece-wise linear fit. a) The variation of drift rates (degrees/year) with solar cycle number for both northern and southern hemispheres in latitude LPZ ranging from $55^\circ$ to a separator latitude from which there is change in linear fit; b) the variation of drift rates (both in degrees/year and m/s) with solar cycle number for both northern and southern hemispheres in latitude HPZ ranging from separator latitude to $\approx 85^\circ$. It can be observed  that the piece-wise linear fits successfully depicts cycle-to-cycle variation in drift rates along with north-south asymmetry taking in account the error bars. For both the latitude zones cycle 19 shows highest drift rate in northern hemisphere and it dominates south for latitude LPZ.}
\label{drift_l}
\end{figure}

The results on migration rates (units both in $^{\circ}$/year and m/s) after applying piece-wise linear fit method are tabulated in Table~\ref{rate_det2} and Figures~\ref{drift_l}a, b. At LPZ the rates were seen to vary between  2.0$\pm$0.1 m/s and 10.1$\pm$1.0 m/s (rate values within brackets in Table~\ref{rate_det2} and right vertical axis in Figure~\ref{drift_l}a). Clear increase of migration rate was seen in HPZ on an average as the rates were seen to vary between 1.3$\pm$0.1 m/s and 19.5$\pm$5.7 m/s (Table~\ref{rate_det2} and Figure~\ref{drift_l}b). Cycle 19 was seen to dominate the migration rates in LPZ north and south. Also, rate in HPZ north was seen to be dominated by cycle 19. Because of higher error bars, same cannot be said for HPZ south. The average separator latitudes automatically found by the piece-wise linear fit came to be $67.5^\circ$ and $69.2^\circ$ for north and south respectively. 

\subsection{Correlation with Available Catalogues}
\begin{figure}
\centerline{
\includegraphics[scale=0.4]{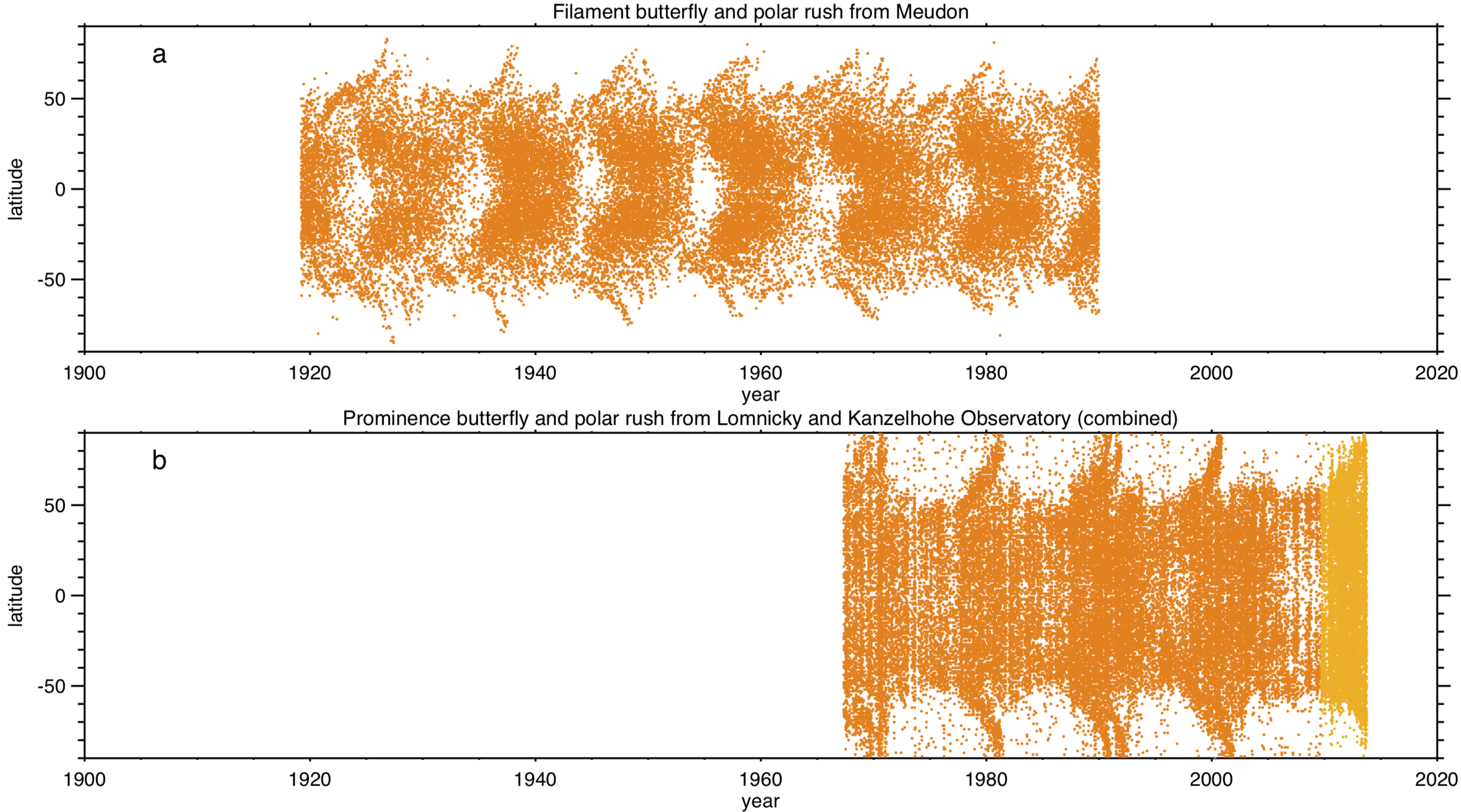}}
\caption{Available filament and prominence catalogues. a) Time-latitude distribution of on-disc filament centers from Meudon H$_{\alpha}$ synoptic map catalogue; b) Time-latitude distribution of H$_{\alpha}$ prominences from Lomnicky (orange) and Kanzelhohe (yellow) catalogue.}
\label{rush_compare}
\end{figure}

\begin{figure}
\centerline{
\includegraphics[scale=0.65]{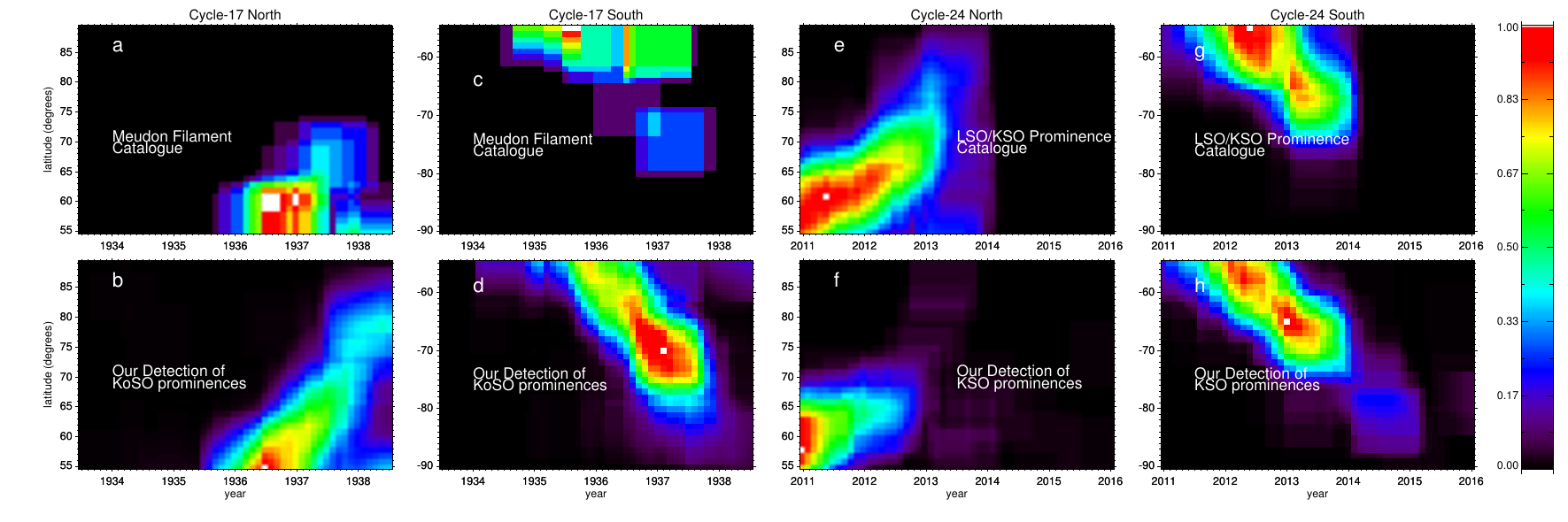}}
\caption{Comparison of polar rush between available catalogues and our detection. a) Density image of cycle-17 northern polar branch derived from Meudon filament catalogue; b) Density image of cycle-17 northern polar branch derived from our detection of KoSO prominences; c) Density image of cycle-17 southern polar branch derived from Meudon filament catalogue; d) Density image of cycle-17 southern polar branch derived from our detection of KoSO prominences; e) Density image of cycle-24 northern polar branch derived from LSO/KSO prominence catalogue; f) Density image of cycle-24 northern polar branch derived from our detection of KSO prominences; g) Density image of cycle-24 southern polar branch derived from LSO/KSO prominence catalogue; h) Density image of cycle-24 southern polar branch derived from our detection of KSO prominences. A colobar has been provided on the right to compare the normalised prominence densities from different observatories.}
\label{pole_compare}
\end{figure}

To validate our prominence detection and understand the effect of using two different wavelengths on statistical basis, we compared our results with the available filament and prominence catalogues from Meudon H$_\alpha$ synoptic maps and LSO/KSO cross calibrated \citep{2011CoSka..41..133R} H$_\alpha$ series respectively (Figures~\ref{rush_compare}a,b). To accomplish this, we first converted the butterfly diagram of our detection and also available catalogues to 2-dimensional histograms with temporal binning of 0.1 year and spatial binning of 1$^\circ$ latitude. As our results primarily focus on polar-rush, we segregated the polar branches satisfying $|latitude|>55^\circ$. For every cycle, we smoothed those polar branches with a kernel of dimension 1 year$\times$ 10$^\circ$ (Figures~\ref{pole_compare}a-h). For every cycle, we correlated those smoothed histogram density images from our detection and the catalogues. The results of correlation for both north and south analysis are presented in Table~\ref{pole_corr}. It can be observed that all the correlations vary within a range of $\approx0.6-0.9$. It can be observed from the histogram images (Figures~\ref{pole_compare}a-d) that for older cycles, where KSO data is not available, KoSO prominence data have an advantage over Meudon filament catalogue in depicting polar rush to higher latitudes with better data density and continuity. Poor data density in polar branch is also responsible for lower correlation coefficient.
The visual match between our detection and LSO/KSO catalogue for cycle-24 (Figures~\ref{pole_compare}e-h) is also reflected in correlation coefficients ($>$0.6) of Table~\ref{pole_corr}.
 

\begin{table}
\begin{center}
\caption{Pole-ward migration rates at different latitudes through piece-wise linear fit}\label{rate_det2}
\resizebox{\columnwidth}{!}{%
\begin{tabular}{ c  c  c  c  c  c  c  c  c  c  c  c  c  c  c}
\hline
Observatory  & Cycle &&&rate  in $^{\circ}$/year (m/s) &&\vline&   separator latitude \\
&& LPZN   &  HPZN  & LPZS   &  HPZS  &\vline&  N & S\\	   \hline  
 	   \hline  
 	  Kodaikanal Ca~{\sc ii}~K &  15 &  8.7$\pm$0.8 & 10.4$\pm$0.4 & 9.2$\pm$0.5 & 20.6$\pm$2.8 &\vline& 65$^\circ$ & 71$^\circ$\\
	  && (3.4$\pm$0.3) & (4.0$\pm$0.2) & (3.6$\pm$0.2) & (7.9$\pm$1.1) &\vline  \\
	 & 16 & 5.6$\pm$0.5 & 19.3$\pm$1.5 & 8.1$\pm$0.4 & 50.6$\pm$14.8 &\vline& 63$^\circ$ & 71$^\circ$\\ 
	 &&(2.1$\pm$0.2) & (7.4$\pm$0.6) & (3.1$\pm$0.2) & (19.5$\pm$5.7) & \vline\\
	  & 17 &  9.3$\pm$1.0 & 19.7$\pm$1.3 & 11.5$\pm$0.6 & 12.2$\pm$0.5 &\vline& 64$^\circ$ & 70$^\circ$\\ 
	  &&  (3.6$\pm$0.4) & (7.6$\pm$0.5) & (4.4$\pm$0.2) & (4.7$\pm$0.2) & \vline\\
	  &  18 & 10.8$\pm$0.8 & 19.6$\pm$3.2 & 11.0$\pm$1.0 & 23.6$\pm$2.6 &\vline& 70$^\circ$ & 67$^\circ$\\ 
	  &&(4.2$\pm$0.3) & (7.6$\pm$1.2) & (4.2$\pm$0.4) & (9.1$\pm$1.0) &\vline\\
	  & 19 & 13.5$\pm$1.4 & 31.8$\pm$7.5 & 26.2$\pm$2.6 & 40.6$\pm$10.6 &\vline& 72$^\circ$ & 71$^\circ$\\ 
	  && (5.2$\pm$0.5) & (12.3$\pm$2.9) & (10.1$\pm$1.0) & (15.6$\pm$4.1) & \vline\\
	  & 20 & 9.6$\pm$0.6 & 3.4$\pm$0.2 & 7.5$\pm$0.5 & 14.1$\pm$2.2 &\vline& 71$^\circ$ & 68$^\circ$\\ 
	  && (3.7$\pm$0.2) & (1.3$\pm$0.1) & (2.9$\pm$0.2) & (5.4$\pm$0.8) & \vline\\
	  & 21 & 9.3$\pm$0.6 & 22.1$\pm$2.2 & 9.0$\pm$0.3 & 37.5$\pm$12.4 &\vline& 69$^\circ$ & 74$^\circ$\\ 
	  && (3.6$\pm$0.2) & (8.5$\pm$0.8) & (3.4$\pm$0.1) & (14.4$\pm$4.8) &\vline\\
	  &  22 & 11.3$\pm$0.6 & 21.4$\pm$2.4 &&&\vline&69$^\circ$\\ 
	  && (4.4$\pm$0.2) & (8.2$\pm$0.9) &&&\vline\\ \hline 
    Meudon H$_\alpha$ &22 &&&5.6$\pm$0.2 & 16.6$\pm$1.5 &\vline&& 71$^\circ$ & \\ 
    &&&&(2.2$\pm$0.2) & (6.4$\pm$0.6) &\vline \\
    & 23 &  14.1$\pm$1.2 & 18.3$\pm$1.6 & 5.3$\pm$0.3 & 22.5$\pm$2.2 & \vline& 69$^\circ$ & 65$^\circ$\\  
    && (5.4$\pm$0.5) & (7.0$\pm$0.6) & (2.0$\pm$0.1) & (8.7$\pm$0.9) & \vline\\ \hline                  
Kanzelhohe H$_\alpha$ & 24 & 6.7$\pm$0.9 & 11.2$\pm$0.7 & 5.9$\pm$0.5 & 11.7$\pm$0.5 &\vline& 63$^\circ$ & 64$^\circ$\\
&& (2.6$\pm$0.3) & (4.3$\pm$0.3) & (2.3$\pm$0.3) & (4.5$\pm$0.2) &\vline \\ \hline
\end{tabular}
}
\end{center}
\end{table}

\begin{table}[]
\begin{center}
 \caption{Correlation of polar branches from our detection and available catalogues}\label{pole_corr}
\begin{tabular}{llllll}
\hline
Cycle & Our detection vs Meudon catalogue && Our detection vs LSO/KSO catalogue \\ \cline{2-5}
 &  North&  South &  North &  South \\ \hline
16 &  {0.64} &  {0.61} &  {-} &  {-} \\ \hline
17 &  {0.81} &  {0.73} &  {-} &  {-} \\ \hline
18 &  {0.61} &  {0.79} &  {-} &  {-} \\ \hline
19 &  {0.74} &  {0.61} &  {-} &  {-} \\ \hline
20 &  {0.88} &  {0.81} &  {-} &  {-} \\ \hline
21 &  {0.65} &  {0.63} &  {0.79} &  {0.80} \\ \hline
22 &  {0.82} &  {0.61} &  {0.70} &  {0.67} \\ \hline
23 &  {-} &  {-} &  {0.78} &  {0.73} \\ \hline
24 &  {-} &  {-} &  {0.86} &  {0.94} \\ \hline
\end{tabular}
\end{center}
\end{table}

\section{Discussion}\label{Disc}
In this section we will first point out the relevance of our results with earlier results as recorded primarily from H$_{\alpha}$ observations. At KoSO prominences are recorded with  Ca~{\sc ii} K$_3$ line, representing a slightly lower height range in chromosphere as compared to H$_{\alpha}$. Thus one may expect the polar rush as calculated here may have some difference with other catalogues.  However, through the comparison of Meudon  H$_{\alpha}$ filament catalogue and our detected Ca~{\sc ii}~K prominences, a close match in polar branches for several cycles are observed. The signature of filament pole-ward migration in H$_\alpha$ is limited to $\approx 70^\circ$ latitudes as shown by \citet{0004-637X-849-1-44} from KoSO data because of projection effect, poor contrast near limb and artifacts. This effect in also seen in the filament time-latitude distribution presented in \citet{2016SoPh..291.1115T} and \citet{2018ApJ...868...52M}. Thus the detected prominences served as better proxies to identify polar rush till close vicinity of poles.   Clear correspondence between catalogue and detection from same data for overlapping cycles validated our detection technique. Information about prominences from KoSO Ca {\sc ii} K disc-blocked images in that sense acts as a complimentary dataset to the H$_\alpha$ filaments. It is also worth mentioning that in terms of time-latitude distribution of prominences, manual detection from the same prominence dataset of KoSO as reported in \citet{1952Natur.170..156A} shows close agreement with the automated detection as presented here. Because the Sun is mostly unipolar during maximum-to-minimum phase the locations detected during that phase may actually represent chromospheric jets near poles. However, such possibilities can only be decoupled through high cadence observations \citep{1999ApJ...520L..71W}.

Though the fitting of polar branches has been done on combined dataset, each polar branch has major contribution from one of the three observatories. So, in the Table~\ref{rate_det2} each polar branch of a cycle has been listed against one among KoSO, Meudon and KSO. The separator latitudes while doing piece-wise linear fits gave an indication that migration rate of change occurs near about $70^\circ$ latitude. The small error-bars also confirms that indeed drift rates keep on increasing towards poles.

Polar rush has important implication to meridional flow \citep{2002ApJ...577L..53W} and turbulent diffusion \citep{Petrovay2017}. The computed polar migration rates as presented in this article, from the  century-long prominence data may effectively constrain such parameters. For example, meridional flow for a solar cycle in the simulation by \citet{2002ApJ...577L..53W} has been scaled according to the solar activity of the same cycle i.e. higher the activity more is the meridional flow rate.   \citet{2001SoPh..202...11M} showed close correlation of integrated sunspot area in rising phase of a cycle having good correlation with pole-ward migration rate. This may further indicate that a correspondence between rate of polar rush and meridional circulation. However, there are certain discrepancies about latitudinal distribution of meridional flow rate near the poles and some studies predict it to be non-existent beyond $60^\circ$ latitudes \citep{Dikpati_2012}. Thus, there can not be a single linear dependence between pole-ward migration rate and meriodional flow rate.
We want to point that polar rush can be caused by combination of diverse mechanisms such as surface diffusion/flux cancellation mimicking poleward migration of magnetic field and not just a passive advective transport. 

As reported in \citet{2012ApJ...753..146M}, such study about proxy of polar field evolution is important as sunspot area combined with polar field evolution gives a complete picture of the long-term variation of solar hemispheric magnetic field. It has been shown before that polar faculae is a good proxy for polar magnetic field with 11 years average periodicity and also that the faculae count become maximum during solar minima \citep{2008ApJ...680.1553S}.

So, the long-term data of prominences as presented here combined with historical polar faculae data can provide a better picture of polar magnetic activity in terms of reversal, build-up processes and may prove to be complementary to recent works on solar cycle predictions relying on long-term data-driven simulations \citep{2018NatCo...9.5209B}.

\section{Conclusions}\label{concl}
 In this paper, we first presented the calibration and processing  of raw disc-blocked Ca~{\sc ii}~K dataset (1906 -- 2002) from the Kodaikanal Solar Observatory. To improve the statistics of prominences  we included the full disc H$_{\alpha}$ data from Meudon and Kanzelhohe observatory till April, 2018 starting from 1980. Salient points of this work are listed below.

\begin{enumerate} 
\item This study presented an unexplored series of disc-blocked Ca~{\sc ii}~K images for the first time.

\item Disc centered, rotation corrected images and subsequent polar maps of same size with constant disc radius were produced for the entire available dataset.  Using the same automated algorithm, prominence locations were detected in all three datasets. To the best of our knowledge this is probably the first  detailed result presented for century-long prominence observation for about 8 cycles from Kodaikanal and 3 cycles from Meudon and Kanzelhohe. All previous studies which contain long-term prominence detection from early 1900's either considered hand-drawn datasets \citep{2016SoPh..291.1115T} or semi-automated detection methods \citep{2011CoSka..41..133R}.

 \item Time-latitude distribution of detected prominences has been produced depicting a clear signature of polar rush for 10 cycles starting from 1906 till April, 2018 with initial ones mimicking those presented in \citet{1952Natur.170..156A} from same dataset. Cycle 20 was seen to have two northern polar branches with the second one being fainter and having higher migration rate. This is consistent with previous studies \citep{1973SoPh...28..389W,1998SoPh..177..357M}.
  \item  Drift rates of polar prominences with latitude greater than $55^\circ$ has been calculated with error bars through piece-wise fits (Table ~\ref{rate_det2}). Two zones (LPZ and HPZ) of polar-branch, found from the fits, clearly depicted change in migration rate with an increase from LPZ to HPZ for most of the cycles. This implies that the latitudes of polar prominences change non-linearly with time. The polar branches are seen well above 70$^\circ$ latitudes with nonzero migration rates whereas studies suggest meriodional flow rate to vanish well before reaching poles \citep{{Dikpati_2012},{van_Ballegooijen_1998}}. Thus the implication of poleward migration rates on meridonal flow needs to be well understood. 
  
  \item Through piece-wise linear fits we found cycle to cycle variation of rates outside error bars for both the lower (LPZ) and higher (HPZ) latitude zones greater than $55^\circ$ latitude. In LPZ, cycle 19 rate dominated others for both north and south. In HPZ only northern hemisphere was seen to be dominated by cycle 19.
  \item Correlation of polar branches with available filament/prominence catalogues depicted match with correlation $>$0.6, validating our detection with an advantage of KoSO data dating back to the early part of last century.
  
    \end{enumerate}

However, this study is not devoid of limitations which we list below:
\begin{enumerate} 
\item The datasets used in this study have not been cross-calibrated and thus the current images cannot be mixed to produce a single uniform disc-blocked dataset.  That is the reason behind us presenting results from each observatory for non-overlapping periods. 
\item We could not present the accuracy of the detection method or effect of the method on uncertainty in extracted results due to the lack of baseline detection method/manual annotations of prominences.
\end{enumerate}

As a follow up work, we would like to address the above limitations and also combine  KoSO H$_\alpha$ data series with this long-term off-limb Ca~{\sc ii}~K dataset to associate filament morphologies and prominences eruptions. Coronal green line (5303\AA )~data \citep{Petrovay2017,1985BAICz..36Q..61S,1998ASPC..150..484M,2007SoPh..241..263M} and space based datasets will be useful sources to add to KoSO prominence dataset for bringing continuity. We hope that these newly calibrated data, as presented here for the first time will enhance the resources of long term archives.

\acknowledgments
 We acknowledge all the observers at Kodaikanal over 100 years for their contribution to build this enormous resource. The Kodaikanal data archive is hosted at  \url{http://kso.iiap.res.in}. We acknowledge the observers of  Meudon and Lomnicky-Kanzelhohe Observatory who made their data available through \url{http://bass2.obspm.fr/sitemap.php} and \url{https://www.astro.sk/~choc/open/lso_kso_h_alpha_promimence_catalogue/lso_kso_h_alpha_promimence_catalogue.html} respectively. We are grateful to the team members of the project on Reconstructing Solar and Heliospheric Magnetic Field Evolution Over the Past Century supported by the International Space Science Institute (ISSI), Bern, Switzerland for giving important suggestions to improve the manuscript. We would like to acknowledge the IUSSTF/JC-011/2016 project grant for supporting this work. We also thank the Science \& Engineering  Research Board (SERB) for the project grant (EMR/2014/626). 

 \end{document}